\input harvmac
\vskip 1cm

 \Title{ \vbox{\baselineskip12pt\hbox{  Brown Het-1292 }}}
 {\vbox{
\centerline{  Exact Correlators of Giant Gravitons    }
\vskip.08in
\centerline{   from  }
\vskip.08in
\centerline{  Dual  $N=4$ SYM theory }  }}

\centerline{$\quad$ { Steve Corley, Antal Jevicki, Sanjaye Ramgoolam   } }
\smallskip
\centerline{{\sl Department of Physics}}
\centerline{{\sl Brown  University}}
\centerline{{\sl Providence, RI 02912 }}
\smallskip
\centerline{{\tt scorley@het.brown.edu }}
\centerline{{\tt   antal@het.brown.edu }}
\centerline{{ \tt ramgosk@het.brown.edu }}

\vskip .3in 
 
 A class of  correlation functions of half-BPS composite 
 operators are computed exactly ( at finite $N$  ) in 
 the zero coupling limit of $N=4$ SYM theory. 
 These have a simple dependence on the four-dimensional 
 spacetime coordinates and are related to correlators 
 in a one-dimensional Matrix Model with complex Matrices obtained by 
 dimensional reduction of $N=4$ SYM on a three-sphere. 
 A key technical tool is Frobenius-Schur duality between 
 symmetric and Unitary groups   and the results are 
 expressed simply in terms of $U(N)$ group integrals or
  equivalently in terms of Littlewood-Richardson coefficients.
 These correlation functions 
 are used to understand the  existence/properties of giant gravitons 
 and related solutions in the string theory dual on $ AdS_5 \times
 S^5$.  Some of their properties hint at integrability 
 in $N=4$ SYM.

\Date{ November 2001 } 

\lref\mst{ J. Mc Greevy, L. Susskind and N. Toumbas, ``Invasion of the
giant gravitons from ADS space,'' hep-th/0003075, JHEP 0006 (2000) 008 } 
\lref\grta{ D. Gross and W. Taylor, ``Two Dimensional QCD is a String
Theory,''  hep-th/9301068, Nucl.Phys. B400 (1993) 181-210 } 
\lref\grtaii{ D. Gross and W. Taylor, ``Twists and Wilson Loops in the
String Theory of Two Dimensional QCD,'' Nucl.Phys. B403 (1993) 395-452 } 
\lref\sch{ S. Naculich, H. Riggs, H. Schnitzer, ``Two Dimensional Yang
Mills Theories are String Theories, '' hep-th/9305097,
 Mod.Phys.Lett.A8:2223-2236,1993  } 
\lref\onsp{ S. Ramgoolam, `` Comment on Two Dimensional
 $O(N)$ and $Sp(N)$ Yang Mills Theories as String Theories, ''
hep-th/9307085,  Nucl.Phys. B418 (1994) 30-44 } 
\lref\cmr{ S. Cordes, G. Moore,  S. Ramgoolam, ``Large N 2D Yang-Mills
    Theory and Topological String Theory,'' hep-th/9402107, 
    Commun.Math.Phys. 185 (1997) 543-619} 
\lref\cmrii{ S. Cordes, G. Moore, S. Ramgoolam, ``Lectures on 2D
Yang-Mills Theory, Equivariant Cohomology and Topological Field
Theories,''hep-th/9411210, Nucl.Phys.Proc.Suppl. 41 (1995) 184-244 } 
\lref\ori{ O. Ganor, J. Sonnenschein, S. Yankielowicz, 
``The String Theory Approach to Generalized 2D Yang-Mills Theory,'' 
hep-th/9407114, Nucl.Phys. B434 (1995) 139-178. } 
\lref\phor{ P. Horava, ``Topological Strings and QCD in Two
Dimensions,'' hep-th/9311156 } 
\lref\doug{ M. Douglas, `` Conformal Field Theory Techniques in Large
N Yang-Mills Theory,'' hep-th/9311130 } 
\lref\fulhar{ W. Fulton and J. Harris, ``Representation Theory, ''
Springer Verlag, 1991  } 
\lref\fulhari{  W. Fulton and J. Harris, ``Representation Theory, ''
Springer Verlag, 1991, pg. 54 } 
\lref\anti{ Antal Jevicki ``Non-perturbative collective field
theory,'' Nucl. Phys. B 376 (1992) 75-98. } 
\lref\jevand{ Antal Jevicki, A. van Tonder, ``Finite [ Q-oscillator ]
representation of 2-D string theory,'' hepth/9601058.  } 
\lref\hashitz{ A. Hashimoto and N. Itzhaki, 
`` Large branes in AdS and their field theory dual,'' hep-th/0008016, 
JHEP 0008 (2000) 051} 
\lref\bbns{ V. Balasubramanian, M. Berkooz, A. Naqvi, M. Strassler, 
``Giant gravitons in conformal field theory,'' hep-th/0107119.   }  
\lref\lmrs{S. Lee, S. Minwalla, M. Rangamani, and N. Seiberg,
``Three-Point Functions of Chiral Operators in D=4, ${\cal N}$=4
SYM at Large N'', Adv. Theor. Math. Phys. 2 (1998)
697, [hep-th/9806074].}
\lref\ehpsw{B.U. Eden, P.S. Howe, A. Pickering, E. Sokatchev, and
P.C. West, ``Four-point functions in N=2 superconformal field
theories'', Nucl. Phys. {\bf B581} (2000) 523, [hep-th/0001138].}
\lref\ehsw{B.U. Eden, P.S. Howe, E. Sokatchev, and
P.C. West,``Extremal and next-to-extremal n-point correlators
in four-dimensional SCFT'', Phys. Lett. {\bf B494} (2000) 141,
[hep-th/0004102]. }
\lref\baryons{E. Witten, ``Baryons and Branes in Anti de Sitter
Space'', JHEP 9807 (1998) 006, [hep-th/9805112].}
\lref\dvv{ R. Dijkgraaf, E. Verlinde, H. Verlinde, 
``BPS Quantization of the Five-Brane,''  hep-th/9604055, 
 Nucl.Phys. B486 (1997) 77-88 }
\lref\witten{ E. Witten, ``Bound States Of Strings And $p$-Branes,''
 hep-th/9510135,  Nucl.Phys. B460 (1996) 335 } 
\lref\brs{   Micha Berkooz, Moshe Rozali, Nathan Seiberg
 ``Matrix Description of M-theory on $T^4$ and $T^5$,''hep-th/9704089, 
Phys.Lett. B408 (1997) 105-110 }
\lref\seib{ Nathan Seiberg,
`` Matrix Description of M-theory on $T^5$ and $T^5/Z_2$,''
hep-th/9705221, Phys.Lett. B408 (1997) 98-104 }
\lref\malda{ J. Maldacena, `` The large N limit of superconformal 
                             field theories and supergravity,'' 
                      Adv.Theor.Math.Phys.2: 231-252, 1998, 
                       hepth/9711200 }
\lref\gkp {S.S.Gubser,I.R.Klebanov and A.M.Polyakov, 
{``Gauge Theory Correlators from Non-Critical String Theory,''}
hepth-9802109 }
\lref\withol{ E. Witten, 
{``Anti-de-Sitter space and holography,''} hepth-9802150,  \hfill\break
 Adv.Theor. Math.Phys. 2 (1998) 253-291 } 
\lref\malstrom{ J. Maldacena, A. Strominger, ``ADS3 black holes 
                   and a Stringy Exclusion Principle,'' hepth/9804085,
JHEP 9812 (1998) 005   }   
\lref\jevram{  A. Jevicki, S. Ramgoolam, 
``Non commutative gravity from the ADS/CFT
          correspondence,'' hep-th/9902059,  JHEP 9904 (1999) 032 } 
\lref\hrt{  P.M.Ho, S.Ramgoolam and R.Tatar, ``Quantum Space-times 
and Finite N
Effects in 4-D Superyang-mills Theories,'' hep-th/9907145, 
Nucl.Phys.{\bf B573} (2000) 364,
}
\lref\gubs{ S. Gubser, {``Can the effective string see higher partial 
                         waves ?''}
 Phys. Rev. D56: 4984-4993, hepth/9704195 }
\lref\myers{ R. Myers, ``Dielectric-Branes, ''hep-th/9910053, 
  JHEP 9912 (1999) 022.  } 
\lref\djr{ S. Das, A. Jevicki, S. Mathur, 
 ``Vibration modes of giant gravitons,'' hep-th/0009019, 
Phys.Rev. D63 (2001) 024013. } 
\lref\djri{ Sumit R. Das, Antal Jevicki, Samir D. Mathur, 
``Giant Gravitons, BPS bounds and Noncommutativity,'' hep-th/0008088, 
Phys.Rev. D63 (2001) 044001 }
\lref\gmt{  T. Grisaru, Robert C. Myers, Oyvind Tafjord, 
 ``SUSY and Goliath, '' hep-th/0008015, JHEP 0008 (2000) 040 } 
\lref\bv{ M. Berkooz and H. Verlinde, {``Matrix Theory, AdS/CFT and
Higgs-Coulomb Equivalence,'' }  hep-th/9907100, 
JHEP 9911 (1999) 037 } 
\lref\holi{ Pei-Ming Ho, Miao Li, 
``Fuzzy Spheres in AdS/CFT Correspondence and
 Holography from Noncommutativity,'' hep-th/0004072, 
Nucl.Phys. B596 (2001) 259-272. }

\lref\guma{M. Gunaydin, N. Marcus, ``The spectrum of the $S^5$
compactification of the chiral $N=2$ $D=10$ supergravity and the
Unitary supermultiplets of $U(2,2/4)$,'' Class. and Quant. Grav. 2,
L11(1985)  } 
\lref\gumz{M. Gunaydin, D. Minic, M. Zagermann, 
``Novel supermultiplets of su(2,2|4) and the AdS(5) / CFT(4)
duality,'' hep-th/9810226, Nucl.Phys.B544:737-758,1999 }
\lref\ski{ Witold Skiba, ``Correlators of short multitrace 
operators in $N=4$  supersymmetric yang-mills,'' hep-th/9907088,
Phys.Rev.D60:105038,1999 } 
\lref\agmo{ O. Aharony, S. Gubser, J. Maldacena, H. Ooguri, 
``Large N Field Theories, String Theory and Gravity,''  hep-th/9905111
Phys.Rept. 323 (2000) 183-386. } 
\lref\baracrule{ A.O.Barut and R.Raczka, ``Theory of Group
Representations and Applications,'' World Scientific, 1986, chapter 8,
pg. 234 }  
\lref\mikh{ A. Mikhailov ``Giant Gravitons from Holomorphic
Surfaces,''hep-th/0010206,  JHEP 0011 (2000) 027 } 
\lref\dh{ E. D'hoker, A. Rhyzov
``Three-Point Functions of Quarter BPS Operators in N=4 SYM,'' 
hep-th/0109065. } 
\lref\rhy{ A. Rhyzov,  ``Quarter BPS Operators in N=4 SYM,'' 
hep-th/0109064. }  
\lref\dhfr{ E. D'Hoker, D.Z. Freedman, S.D.Mathur, A. Matusis,
L. Rastelli, ``Extremal Correlators in the AdS/CFT Correspondence, '' 
hep-th/9908160 } 
\lref\barac{ A.O.Barut and R.Raczka, ``Theory of Group
Representations and Applications,'' World Scientific, 1986 } 
\lref\af{ L. Andrianopoli, S. Ferrara,
``On short and long SU(2,2/4) multiplets in the AdS/CFT
correspondence,'' hep-th/9812067, Lett.Math.Phys. 48 (1999) 145-161 
}
\lref\afsz{ L. Andrianopoli, S. Ferrara, E. Sokatchev, B. Zupnik
``Shortening of primary operators  in N-extended SCFT-4 
 and harmonic-superspace analyticity,''
 hep-th/9912007, Adv.Theor.Math.Phys. 3 (1999) 1149-1197
 }

\def\nro{ n_{R_1} } 
\def\nrt{ n_{R_2} }

\def\nri{ n_{R_i} } 
\def\nsj{ n_{S_j} } 
\def\rop{ R_1^{\prime} } 
\def\rtp{ R_2^{\prime} } 
\def\sop{ S_1^{\prime} } 
\def\stp{ S_2^{\prime} } 
\def\rlp{ R_l^{\prime} } 
\def\skp{ S_k^{\prime} } 
\def\rip{ R_i^{\prime} } 
\def\sjp{ S_j^{\prime} } 
\def\car{\cal R}

\newsec{ Introduction } 

The AdS/CFT correspondence \refs{\malda, \gkp, \withol }
presents an opportunity to investigate new qualitative 
 features of non-perturbative string theory using 
techniques in the dual conformal field theory. 
Finite $N$ truncations in  BPS spectra 
were studied as evidence of a stringy exclusion 
principle \malstrom ( see also \gubs ). These were argued to provide 
evidence for non-commutative gravity  in 
\jevram, and related developments appeared in \bv\hrt\holi.  
 
 In another line of development the work 
of \myers\ showed that in the presence of background 
fields, zero-branes can get polarized into higher branes. 
Giant gravitons were discovered in \mst, combining 
heuristic expectations from non-commutativity and 
the Myers effect, and exhibiting time-dependent solutions 
to brane actions which describe branes extended in the 
spheres of the $ AdS \times S$ backgrounds. Giant gravitons 
were further studied in \refs{ \djr, \djri, \gmt, \hashitz }. 
One of the results of these papers was that there are new 
giant gravitons which are large in the $AdS $ directions.  
A number of puzzles were raised about the detailed 
 correspondence between the spectrum of chiral primaries 
 and the spectrum of  gravitons :  Kaluza-Klein gravitons, 
 sphere giants and AdS giants. Some of these puzzles were addressed
recently in  \bbns\ and the gauge theory dual 
of a sphere giant was proposed.

 In this paper we will develop the dictionary 
 between half-BPS operators and Yang Mills. 
 We will observe that the space of half-BPS representations
 can be mapped to the space of Schur 
 polynomials of $U(N)$, equivalently to the space of Young Diagrams
 characterizing representations of $U(N)$. 
 We will compute correlation functions involving arbitrary 
 Young diagrams. The map to Young Diagrams will reveal natural 
 candidates for sphere giants, agreeing with \bbns, 
 as well as candidates for AdS giants. In addition we will 
 be lead to look for a simple generalization of sphere and 
 AdS giant graviton solutions involving multiple windings. 
 We will discuss  the properties of the correlation 
 functions in the light of the correspondence to giant 
 gravitons.

\newsec{ Review and Notation.  } 

We will recall some facts about half-BPS operators, 
 free field contractions, 
 symmetric groups and Unitary groups. The connection between 
 Symmetric and Unitary groups was crucial in the development 
 of the String theory of two-dimensional Yang Mills Theory 
\refs{ \grta, \grtaii,\sch,  \onsp , \cmr \ori, \doug, \phor } 
 and in certain aspects of Low-dimensional Matrix Models \refs{ \anti, 
 \jevand }.  
 Some key useful results are collected here. Many  of the 
 key derivations are reviewed in \cmrii. 

\subsec{ Half-BPS operators in $N=4 $ SYM } 

We will review some properties of 
half-BPS operators in $N=4$ SYM \refs{\af, \afsz,  \guma, \gumz, \ski }.
A more complete list of references is in \agmo.

The half-BPS operators constructed 
from the $6$ real scalars  in the Yang-Mills theory lie
in the $(0,l,0)$ representation of the $SU(4) \sim SO(6)$
${\cal R}$-symmetry group. This is the symmetric 
 traceless representation of $SO(6)$ corresponding 
 to Young Diagrams with one row of length $l$. 
  These operators saturate a lower
bound on their conformal dimensions which is related to
their ${\cal R}$-symmetry charge.
They include single trace as 
 well as multiple trace operators. 
The single trace chiral
primary operators are of the form
$T_{i_1 \cdots i_6} \, Tr(X^{i_1} \cdots X^{i_n})$ where the $X^i$'s, with
$i=1,...,6$, are the
scalars of the theory transforming in the vector
representation of the $SO(6)$ ${\cal R}$-symmetry group 
and the coefficients $T_{i_1 \cdots i_6}$ are symmetric
and traceless in the $SO(6)$ indices.

The ${\cal N}$=4 theory can be decomposed into
an ${\cal N}$=1 vector multiplet and three chiral multiplets.
The scalars of the chiral multiplets are given in terms of
the $X^j$'s as $\Phi^j = X^j + i X^{j+3}$ for $j=1,2,3$ where
all fields transform in the adjoint.  In this notation the
$SO(6)$ ${\cal R}$-symmetry is partially hidden so that only
a $U(3)$ action remains explicit. 
The $U(3)$ decomposition of the 
 rank $l$ symmetric traceless tensor  representation of $SO(6) $ 
 includes one copy of the rank $l$ symmetric tensor representation 
 of $U(3)$.  
Single trace operators of this form are  of the form 
$\alpha_{j_1 \cdots j_l} Tr(\Phi^{j_1} \cdots \Phi^{j_l})$, 
 where  $\alpha$ is symmetric
in the $U(3)$ indices, without the need for tracelessness.
In this  class of chiral primaries 
$Tr((\Phi^1)^l)$ appears once.
Each single trace half-BPS   representation contains 
precisely one such operator. 
We will henceforth drop the $1$ and write this 
as $Tr((\Phi)^l)$, with the understanding that we have 
fixed a $U(3)$ index.

The argument can be generalized to multi-trace 
operators. A two-trace composite of scalars 
will take the form $T_{i_1 \cdots i_{l_1} i_{l_1+1}  
\cdots i_{l_2} } tr X^{i_1}  \cdots X^{i_{l_1} } tr X^{i_{l_1+1}}
\cdots X^{i_{l_2}} $, where $T$ is a traceless symmetric
 tensor of $SO(6)$. This uses the fact that the 
correspondence between    traceless symmetric reps ( i.e of type 
$(0,l,0)$ ) and the half-BPS condition holds irrespectively
of how we choose to contract the gauge indices, as emphasized for 
 example in \ski\dhfr. Using again the fact that the 
traceless symmetric tensor decomposes into $U(3)$ 
 representations which include the symmetric representation 
 built from the $l_1+l_2$-fold tensor product, we 
know that $ tr ( \Phi^{l_1} ) tr ( \Phi^{l_2} ) $ will 
 appear once  in this set of operators. 
While it is obvious that $tr ( \Phi^{l_1} ) tr ( \Phi^{l_2} ) $
preserves the same supersymmetry as $ tr ( \Phi^{l_1} )$, we have 
now shown that every double trace half-BPS 
 representation contains one operator of this form. 
By a similar argument,
the multi-trace chiral primaries will include 
$Tr( \Phi^{l_1})Tr ( \Phi^{l_2} ) \cdots $. 
More generally for a fixed  \car\ charge of $n$ 
there will be operators of the form 
\eqn\genfrm{ 
( tr ( \Phi^{l_1} ) )^{k_1} tr  ( \Phi^{l_2} ) )^{k_2} \cdots 
( tr  ( \Phi^{l_m} ) )^{k_m} } 
where the integers $l_i,k_i $ form a partition 
of $n$ 
\eqn\partn{ n = \sum_{i=1}^{m} l_i k_i }  
We have, therefore,  a one-to-one correspondence then between half-BPS 
 representations  of charge $n$ and partitions 
of $n$.  A useful basis in this space of operators 
is given by Schur Polynomials of degree $n$ 
for the unitary 
group $U(N)$.  We will return to 
these in section 2.4. 

We will associate a Schur polynomial in the complex matrix $ \Phi $ 
to each short representation and we will compute  correlators 
which are obtained by considering overlaps of such holomorphic
polynomials and their conjugates involving $ \Phi^{*}$. 
Each observable by itself does not contain both 
$ \Phi$ and $ \Phi^{*}$. 
 This gives us 
a special relation between the weights of the 
operators involving $\Phi$ and those involving $\Phi^{*}$. 
These special correlators are called extremal correlators
in the literature. There exists a
non-renormalization theorem \refs{\ehpsw, \ehsw } 
protecting extremal correlators 
of the half-BPS chiral primaries ( single and multi-trace  ) 
  so that the weak coupling
computation of the correlators can be safely extrapolated
to strong coupling without change. 
To compute more general correlation functions, one would
need $ SO(6)$ descendants of $tr ( \Phi^l ) $ ( and its multi-trace 
generalizations )  which 
will typically involve all three complex $\Phi$'s and 
their conjugates.

\subsec{ Free fields and Combinatorics } 

The basic two-point function we will 
need follows from the free field correlator : 
\eqn\frfi{ < \Phi_{ij}(x_1)  \Phi^{*}_{kl}(x_2)  > 
= {\delta_{ik} \delta_{jl} \over (x_1 - x_2 )^2 } }
Our main focus will be on the dependence of 
the correlators on the structure of the composites
and $N$, so we will often suppress the 
spacetime dependences.  

Free field contractions in $U(N) $ 
gauge theory will frequently lead to sums 
of the form 
\eqn\manyind{ 
\sum_{i_1 , i_2 \cdots i_n } \delta^{i_1}_{i_{\sigma(1)} }  
\delta^{i_2}_{i_{\sigma(2)} } \cdots \delta^{i_n}_{i_{\sigma(n)} } }
where each index $i_1, \cdots i_n$ runs over integers from 
$1$ to $N$. If $\sigma $ is the identity then the above sum 
is $N^n$. If $\sigma$ is a permutation with one cycle of 
length $2$ and remaining cycles of length $1$, then it is 
$N^{(n-1)}$. One checks that more generally the sum is $N^{C(\sigma)} $
 where $ C ( \sigma )$ is the number of cycles 
in the permutation $ \sigma$. 
\eqn\maninda{ 
\sum_{i_1 , i_2 \cdots i_n } \delta^{i_1}_{i_{\sigma(1)} }  
\delta^{i_2}_{i_{\sigma(2)} } \cdots \delta^{i_n}_{i_{\sigma(n)} } 
 = N^{ C(\sigma) } } 

Rather than writing out strings of delta functions
carrying $n$ different indices, we will use a multi-index 
$I(n) $ which is shorthand for a set of $n$ indices.
We will also use $I(\sigma(n)) $ for a set of $n$ indices 
with their labels shuffled by a permutation $\sigma$. With 
this notation, \maninda\ takes the form
\eqn\idpow{
 \sum_{I} \delta \pmatrix{ & I(n) \cr 
                  & I ( \sigma (n)  ) \cr } 
 = N^{C( \sigma ) } } 
We will often need a slight variation of this result,
\eqn\idpowi{
 \sum_{I} \delta \pmatrix{ & I( \alpha (n))  \cr 
                  & I ( \beta (n)  ) \cr } 
 = N^{C( \beta \alpha^{-1} ) } = N^{C (\alpha^{-1} \beta  }) } 
which follows trivially from the previous relation after
noting that the left-hand-side is invariant under
the replacement $n \rightarrow \alpha^{-1} (n)$, or by 
acting on the pair $ (\alpha, \beta ) $ with 
$ \alpha^{-1} $ from the left, to reduce the LHS of 
\idpowi\ to that of  \idpow.

\subsec{ Symmetric groups }

After  performing  the sums over contractions we will obtain
sums over permutations, or equivalently 
over elements of permutation groups. It will be useful 
to manipulate quantities which are formal 
sums over symmetric group elements. These live in the group 
algebra of $S_n$. 
An interesting function on the group algebra is the delta function 
which is $1$ when the argument is the identity and $0$ otherwise. 
This function has an expansion in characters
\eqn\deltaeq{ \delta ( \rho ) = { 1 \over n! } \sum_R d_R ~~ \chi_R
( \rho ) } 
The sum is over representations $R$  of $S_n$ which are associated with 
Young Diagrams with $n$ boxes. $d_R$ is the dimension of a 
representation $R$ and $\chi_R(\rho)$ the character, or trace,
of the element $\rho \in S_n$ in the representation $R$. 

Another useful fact is that characters, in an irreducible
representation of the symmetric group $R$, of a product of 
an element $ \cal C $ of the group algebra which commutes with
everything with an arbitrary element $\sigma$ can be factorized 
as follows : 
\eqn\commt{ 
 \chi_R ( { \cal C } ~ \sigma ) = 
{ \chi_R ( { \cal C } )~  \chi_R ( \sigma)  \over d_R }.   }
Elements $ \cal C $ that we will run into involve 
either averages over the symmetric group
of the form $ \sum_{\alpha, \rho } f( \alpha \rho \alpha^{-1}  ) \rho
$, or $ \sum_{\rho} g( \rho ) \rho $ where $ g(\rho)  $ is a  
class function. 

We will also need certain orthogonality 
properties. Consider 
\eqn\orth{  \sum_{\sigma } \chi_R ( \sigma^{-1}  ) D_S ( \sigma ) } 
where $D_S ( \sigma ) $ is the matrix representing 
$ \sigma $ in the irreducible representation $S$. 
The matrix written down in \orth\ can  be proved 
to commute with any permutation  $\tau $ acting in the  
representation $S$. By Schur's Lemma it is therefore  proportional 
to the identity. Hence 
\eqn\orthi{  \sum_{\sigma } 
\chi_R ( \sigma^{-1}  ) D_S ( \sigma ~ \alpha  )  =
\sum_{\sigma }  \chi_R ( \sigma^{-1}  )~ { \chi_S ( \sigma  ) \over d_S
}  ~ D_S( \alpha)    } 
Now we can use the orthogonality of characters 
\eqn\orthii{ 
\sum_{\sigma } 
\chi_R ( \sigma^{-1}  ) ~ {\chi_S ( \sigma  ) }  =
  \delta_{RS } ~ n!  } 
in order to simplify \orthi\ 
\eqn\oh{ \sum_{\sigma } 
\chi_R ( \sigma^{-1}  ) ~ D_S ( \sigma \alpha  ) 
= { \delta_{RS} ~ n ! \over d_S }  D_S ( \alpha ) } 
Taking a trace in the representation $S$ we get 
\eqn\ohi{ \sum_{\sigma } 
\chi_R ( \sigma^{-1}  ) ~ \chi_S ( \sigma \alpha  ) 
= { \delta_{RS} ~ n ! \over d_S }  \chi_S ( \alpha ) }

\subsec{ Duality between Symmetric Groups and Unitary Groups } 

Denote by $V$ the fundamental representation of 
$U(N)$. The space $ Sym ( V^{ \otimes n } ) $ is a representation 
of $U(N)$ and also admits a commuting action of 
 $S_n$. The action of $U(N)$ and $S_n$ can thus be simultaneously 
diagonalised. This allows Young Diagrams to be associated with 
both $U(N)$ and $S_n$ representations. Some results following from 
this connection are summarized in the following. More details can be
found in \refs{ \fulhar \barac}  for example. 

The Schur polynomials are characters of the unitary 
group in their irreducible representations. 
\eqn\charu{ \chi_{R} ( U ) = { 1 \over n! } \sum_{ \sigma \in S_n } \chi_{R}
( \sigma)~  tr ( \sigma ~ U ) } 
In the RHS the trace is being taken in 
$V^{\otimes n } $ and both $ \sigma $ and $U$ are operators 
acting on this space. The action of $\sigma $ is given 
by : 
\eqn\actsig{ \sigma  ( v_{i_1} \otimes v_{i_2} \cdots \otimes v_{i_n}  ) 
 =  ( v_{i_{\sigma(1)} } \otimes v_{i_{\sigma(2)}} \cdots
\otimes v_{i_{\sigma(n)}}  ) } 
By inserting $U=1$ we find a formula for the 
dimension of a representation of the unitary group : 
\eqn\shudim{ Dim_N ( R ) = {1 \over n!} \sum_{\sigma \in S_n} 
\chi_{R } ( \sigma ) N^{ C(\sigma ) } } 

We will find it convenient, in this paper, to consider 
the extension of the Schur polynomials from Unitary to Complex 
Matrices, i.e 
\eqn\charp{ \chi_{R} ( \Phi ) = { 1 \over n! } \sum_{ \sigma \in S_n } \chi_{R}
( \sigma) tr ( \sigma \Phi ) } 
These form a basis in the space of $U(N)$ invariant functions 
of the Matrix $ \Phi$ where $ U(N)$ acts on $\Phi $ by conjugation.

The multi-index notation introduced in section 
2.1 is also useful in giving 
compact expressions for $tr ( \sigma \Phi ) $ 
\eqn\cptts{\eqalign{ 
  tr ( \sigma \Phi ) & = 
\sum_{i_1,i_2,  \cdots, i_n } \Phi^{i_1}_{i_{\sigma(1)}}
\Phi^{i_2}_{i_{\sigma(2)}}\cdots  \Phi^{i_n}_{i_{\sigma(n)}} \cr 
 &= \sum_{I} \Phi \pmatrix{ & I(n)~ \cr 
                  & I( \sigma (n) ) ~\cr  } \cr }}

The fusion coefficients of $U(N)$ also have meaning 
in the context of symmetric groups. 
Let $g(R_1,R_2; S)$ be the multiplicity of the 
representation $S$ in the tensor product of representations 
$R_1$ and $R_2$. Let $\nro$, $\nrt$, and $n_S$ be the number of 
boxes in the Young diagrams $R_1$, $R_2$, and $S$ respectively.
$S_{n_{S}} $ contains the 
product $S_{ \nro } \times S_{\nrt }  $ as a subgroup. 
As such the character in $S$ of any permutation which takes 
the form $ \sigma_1 \circ \sigma_2$, a product of two permutations 
where the first acts non-trivially on the first 
$\nro $ elements and the second acts on the last $\nrt $ elements 
of $n_S$, can be decomposed into a product of characters : 
 \eqn\keq{ 
\chi_{S} ( \sigma_1 \circ \sigma_2 ) 
 = \sum_{ R_1 \in Rep (~ S_{n_1} ~ ) }  
    \sum_{  R_2 \in Rep (~ S_{n_2} ~ ) }
    g(R_1, R_2; S ) ~  \chi_{R_1} ( \sigma_1 ) ~ \chi_{R_2 }  ( \sigma_2 )  }
We know such an expansion in characters of the product 
group must exist. It is a non-trivial result from 
the theory of symmetric and unitary groups, related to 
Frobenius-Schur duality, that the coefficients  $g(R_1, R_2, S )$ 
appearing in the 
expansion are the multiplicities with which the representations 
$S$ of $U(N)$ appear in the tensor product of the representations 
$R_1$ and $R_2$ \fulhari. These coefficients can be written in terms of 
$U(N)$ group integrals, and can also be computed using 
 a combinatoric algorithm called the Littlewood-Richardson rule. 

\newsec{ General Results } 

 For two point functions \foot{ We are expressing the results
 in terms of overlaps of $\chi ( \Phi )$ and $\chi ( \Phi^* ) $, 
 but we could replace  the latter with  $\chi ( \Phi^{\dagger}  ) $ 
 which would change some intermediate steps in the derivations, 
 but not the final answers }, we find  
\eqn\twpt{ < \chi_R ( \Phi ) \chi_S ( \Phi^{*} )  > =
  \delta_{RS} { Dim_{N} ( R ) n_{R}! \over d_R} } 
In the above we have suppressed the $x$ dependence
 but they can be restored by conformal invariance 
\eqn\twptx{ < \chi_R ( \Phi ) ( x_1 ) \chi_S ( \Phi^{*} )( x_2 )   > =
  \delta_{RS} { Dim_{N} ( R ) n_R! \over d_R}  { 1 \over (x_1-x_2)^{2n_R} } } 
For three point functions we have 
\eqn\threept{ <  \chi_R ( \Phi )  ~\chi_S ( \Phi )  ~ \chi_T ( \Phi^
  {*}  ) > = g(R,S;T)  
{ n_T! ~Dim_N (T) \over d_T } } 
Restoring the dependence on four dimensional 
space-time coordinates : 
\eqn\threeptx{\eqalign{ 
&  <  \chi_R ( \Phi ) (x_1) ~\chi_S ( \Phi )(x_2) 
  ~ \chi_T ( \Phi^
  {*}  ) (x_3)  > \cr 
& = g(R,S;T)  { n_T!  ~Dim_N (T) \over d_T }
 { 1\over (x_1 -x_3)^{2n_R} (x_2 -x_3)^{2n_S} 
 } \cr }} 

 For $ l \rightarrow 1 $ we have 
 \eqn\nrhto{   < \chi_{R_1} ( \Phi ) \chi_{R_2}  ( \Phi ) \cdots 
  \chi_{R_n}  ( \Phi  )  \chi_{S} ( \Phi^{*}  )  >
  ~= ~ g(R_1, R_2, \cdots R_n ; S ) 
{ n_S! Dim_N ( S) \over d_S }  } 
 Restoring spacetime coordinates :  
\eqn\nrhtox{\eqalign{ 
&    < \chi_{R_1} ( \Phi )(x_1)  \chi_{R_2}  ( \Phi )(x_2)  \cdots 
  \chi_{R_n}  ( \Phi  ) ( x_l)  \chi_{S} ( \Phi^{*}  ) (y)  >
 \cr 
&   = ~ g(R_1, R_2, \cdots R_n ; S ) 
{  n_S! ~ Dim_N ( S) \over d_S }   
 { 1 \over (x_1-y)^{2n_{R_1}}  (x_2-y)^{2n_{R_2}} \cdots
 (x_l-y)^{2n_{R_l}} } \cr }}
 For correlators of this form where the conformal 
 weights are related as above $ n_S = n_{R_1} + \cdots n_{R_l} $, 
 the kind of contractions we have described above are the most general.

 For $ l \rightarrow k $ we have 
 \eqn\nrhtm{\eqalign{ 
&   < \chi_{ R_1 } ( \Phi ) \chi_{R_2}  ( \Phi ) \cdots 
  \chi_{R_l}  ( \Phi  ) ~  \chi_{S_1} ( \Phi^{*}  ) \cdots
\chi_{S_k} ( \Phi^{*} )  >  \cr 
   &= \sum_{S} g(R_1, R_2 \cdots R_l ; S ) 
 {  n_S! ~
Dim_N (S)  \over d_S }  g(S_1, S_2,
\cdots S_k ; S ) }}  
 In this case we are capturing only a special 
 case of the $l \rightarrow k $ multipoint function, 
 where the spacetime coordinates of all the operators 
 involving $ \Phi^{*}$ coincide,
  \eqn\nrhtmx{\eqalign{ 
&   < \chi_{ R_1 } ( \Phi ) (x_1) \chi_{R_2}  ( \Phi ) (x_2) \cdots 
  \chi_{R_n}  ( \Phi  )(x_l )  ~  \chi_{S_1} ( \Phi^{*}  ) (y) \cdots
\chi_{S_m} ( \Phi^{*} )(y)   >  \cr 
   &= \sum_{S} g(R_1, R_2 \cdots R_n ; S ) 
 {  n_S! Dim_N  S \over d_S } g(S_1, S_2,
\cdots S_m ; S )   { 1 \over 
 ( x_1-y)^{2n_{ R_1}} \cdots  (x_l-y)^{2n_{R_k} } } }}

The coefficient $g(R_1, R_2 \cdots R_n ; S )$ is a positive 
integer. 
It can also be expressed in terms of 
Unitary group integrals : 
\eqn\gunint{ g(R_1, R_2, \cdots R_n ; S ) = 
\int dU \bigl( \prod_{i=1}^{n}  \chi_{R_i} ( U) \bigr )  \chi_S
( U^{ \dagger } ) } 
 It  is obtained, in the case of $n=2$ by the 
Littlewood-Richardson rule for fusing Young diagrams ( 
see for example \baracrule). The product $\chi_{R_1} ( U ) \chi_{R_2 } ( U )$ 
can be expanded as $ \sum_{R} g(R_1, R_2; R ) \chi_R ( U ) $. 
Repeated use of this expansion  in the integral \gunint\ leads to 
\eqn\gmgs{ g( R_1, R_2 \cdots R_n; S ) = \sum_{S_1, S_2 \cdots,
S_{n-2}  } 
 g(R_1,R_2; S_1) g(S_1, R_3; S_2) \cdots g(S_{n-2}, R_n;  S ). }

\newsec{ Two point functions } 

 We will derive here formulae for 
 the correlator $ < \chi_R ( \Phi ) \chi_S ( \Phi^{*} )  >$. 
 It follows trivially that this will only be non-zero when 
 the number of boxes in $R$ is the same as the number 
 of boxes in $S$. So we have $n_R = n_S = n$.  
 A more non-trivial fact is that 
 the exact finite $N$ correlator is proportional 
 to $\delta_{RS}$, i.e., it is non-zero only 
 if $R$ and $S$ are the same Young Diagram. 

We first convert the free field computation 
to sums over the symmetric group $S_n$. 
\eqn\twptc{\eqalign{ 
&< \chi_R ( \Phi ) \chi_S ( \Phi^{*} )  > \cr 
 & = < \sum_{\sigma} { \chi_R ( \sigma )\over n ! }  ~~tr ( \sigma \Phi ) 
\sum_{ \tau } { \chi_S (\tau )\over n! } 
 ~~tr ( \tau \Phi^{*} ) > \cr   
 & = \sum_{ i_1, i_2 \cdots i_n } ~~ \sum_{   j_1, j_2 \cdots j_n  } 
\sum_{\sigma} { \chi_R ( \sigma ) \over n! } 
 \sum_{ \tau } { \chi_S (\tau ) \over n! } \cr 
& \qquad\qquad\qquad
  < \Phi^{i_1}_{i_{\sigma(1) }} \Phi^{i_2}_{i_{\sigma(2)}} \cdots 
  \Phi^{i_n}_{ i_{\sigma(n) }} ( \Phi^{*})^{j_1}_{j_{\tau(1) }} 
    ( \Phi^{*})^{j_2}_{j_{\tau(2) }}\cdots   
 ( \Phi^{*})^{j_n}_{j_{\tau(n) }} > \cr 
 &= \sum_{ i_1, i_2 \cdots i_n } ~~ \sum_{   j_1, j_2 \cdots j_n  } 
\sum_{\alpha} \sum_{\sigma, \tau } { \chi_R ( \sigma )\over n!} 
  { \chi_S (\tau  ) \over n! }  ~~ \cr 
& \qquad\qquad\qquad
\delta^{i_1}_{j_{\alpha \tau(1) } } \delta^{i_2}_{j_{\alpha \tau(2) } } 
  \cdots \delta^{i_n}_{j_{\alpha \tau(n)  }} ~~~~  
 \delta^{j_1}_{i_{\alpha^{-1} \sigma(1) }} 
\delta^{j_2}_{i_{\alpha^{-1} \sigma(2) } } \cdots 
\delta^{j_n}_{i_{\alpha^{-1} \sigma(n) } }\cr                  
&=    \sum_{ i_1, i_2 \cdots i_n }~~  
\sum_{\alpha}\sum_{\sigma} { \chi_R ( \sigma ) \over n!} 
 \sum_{ \tau } { \chi_S (\tau ) \over n!} ~~ 
\delta_{i_{\alpha^{-1} \sigma(1) }}^{i_{\tau^{-1} \alpha^{-1}(1)}} \cdots 
  \delta_{i_{\alpha^{-1} \sigma(n) }}^{i_{\tau^{-1} \alpha^{-1}(n)}} \cr 
&=  \sum_{ i_1, i_2 \cdots i_n }
\sum_{\alpha} \sum_{\sigma} { \chi_R ( \sigma )\over n!} 
 \sum_{ \tau } { \chi_S (\tau ) \over n! }   ~~
   \delta_{i_1}^{i_{\sigma^{-1} \alpha\tau^{-1} \alpha^{-1} (1)} } 
\cdots    \delta_{i_n}^{i_{\sigma^{-1} \alpha\tau^{-1} \alpha^{-1} (n)
} } \cr
&=   \sum_{\alpha} \sum_{\sigma} { \chi_R ( \sigma )\over n!} 
 \sum_{ \tau } { \chi_S (\tau )\over n! } ~
N^{C(\sigma^{-1} \alpha \tau^{-1} \alpha^{-1} ) } \cr }}

To show how the notation developed 
in section 2 simplifies the calculations, we redo the same 
maniplulations as above in a slightly more compact form.
For higher point functions we will present the derivations 
exclusively in the more compact form.  
\eqn\twptcpt{\eqalign{ 
& < \sum_{\sigma} { \chi_R ( \sigma ) \over n!} 
 ~~tr ( \sigma \Phi ) \sum_
{ \tau } { \chi_S (\tau ) \over n!}  ~~tr ( \tau \Phi^{* } ) > \cr 
& = \sum_{I,J} ~~ \sum_{\sigma} { \chi_R ( \sigma )\over n!}  \sum_{ \tau } 
{ \chi_S (\tau ) \over n!} 
  < \Phi \pmatrix { & I(n) \cr  
                      &I  (\sigma(n) ) \cr } 
\Phi^{*} \pmatrix { & J(n) \cr  
                      &J (\sigma(n) ) \cr } > \cr 
& = \sum_{I,J} ~~ \sum_{\sigma, \tau, \alpha} { \chi_R ( \sigma ) \over n!} 
 {  \chi_S (\tau ) \over n!  }  
\delta \pmatrix { & I(n) \cr 
                  & J( \alpha \tau (n) ) \cr} 
\delta  \pmatrix { & J(n) \cr 
                  & I( \alpha^{-1}  \tau (n) )  } \cr 
& =  \sum_{I} ~~\sum_{\sigma, \tau , \alpha} 
 {\chi_R ( \sigma ) \over n!} { \chi_S (\tau ) \over n!  }  
\delta \pmatrix { & I(  \tau^{-1} \alpha^{-1}   (n) )  \cr 
                  & I( \alpha^{-1} \sigma (n)  ) \cr} \cr 
& = \sum_{I} ~~ \sum_{\sigma, \tau, \alpha} { \chi_R ( \sigma ) \over n! }  
     { \chi_S (\tau ) \over n! }  
    \delta \pmatrix { &I( \sigma^{-1} 
                     \alpha \tau^{-1} \alpha^{-1} (n) )  \cr 
                      &  I(n) \cr } \cr 
& =  \sum_{\sigma, \tau, \alpha } { \chi_R ( \sigma ) \over n! } 
  { \chi_S (\tau ) \over n! } ~
   N^{C(\sigma^{-1} \alpha \tau^{-1} \alpha^{-1} ) } \cr }} 

Having converted the sum over contractions 
to a sum over symmetric groups we will now 
use the connection between symmetric and Unitary groups
to express the answer in terms of Dimensions of Unitary 
groups.  Introducing a new summed permutation $p$ constrained 
by a delta function, which simplifies the exponent of $N$, 
and then summing over $\tau$ we get, 
\eqn\fanp{\eqalign{  
& \sum_{\alpha,\sigma, \tau, p } { \chi_R ( \sigma )\over n! } ~ 
 { \chi_S (\tau ) \over n!} 
~ N^{C( p   ) } ~ \delta ( p^{-1} 
              \sigma^{-1} \alpha \tau^{-1} \alpha^{-1} )  \cr 
& = {1 \over (n!)^2}  \sum_{\alpha,\sigma,  \gamma }\chi_R ( \sigma ) 
~ \chi_S ( \alpha^{-1} p^{-1} \sigma^{-1} \alpha    )~ 
N^{C( p  ) }  \cr 
& =   { 1 \over n! }  \sum_{ \sigma,  p } 
\chi_R ( \sigma ) ~ \chi_S ( p^{-1} \sigma^{-1} )~  N^{C( p ) } \cr 
& = \sum_{ p } 
{ 1 \over d_R } ~\delta_{RS} ~\chi_{S} ( p^{-1} )~  N^{C( p  ) } \cr 
& =  n! ~ { Dim_N ( R ) \over d_R }  ~  \delta_{RS} \cr }}
In the third line we used cyclic invariance 
of the trace to do the sum over $\alpha$ thus gaining 
a factor $n!$. 
To obtain the   $ { \delta_{RS} \over d_R }$ in the fourth 
line, we used \orthii. In the final step we used \shudim.

The factor $ {n!  Dim ( R ) \over d_R }  $ will come 
up frequently, so we will call it $f_{R}$ and make it 
more explicit. From group theory texts \refs{ \barac, \fulhar }, 
we find the dimensions : 
\eqn\dimdiv{\eqalign{ 
& d_R = { n! \over  \prod_{i,j} h_{i,j} }, \cr 
& Dim_N ( R ) = \prod_{i,j} { (N-i+j) \over h_{i,j} }.   \cr }}
The product runs over the boxes of the 
Young Diagram associated with $R$, with $i$ 
labelling the rows and $j$ labelling the column. 
The quantity $h_{i,j} $ is the hook-length associated
with the box. The quantity $f_R$ takes the simple form: 
\eqn\fr{ f_R = \prod_{i,j}  (N-i+j) }

This orthogonality of operators associated 
with Young Diagrams is reminiscent of 
{ \it Unitary}  Matrix  Models. Indeed if we integrate 
the characters of $U(N)$ as with the unit-normalized 
Haar measure as in $ \int dU \chi_R (U) \chi_S
( U^{\dagger} ) = \delta_{RS}$ we have orthogonality 
with a different normalization factor. 
The result in \fanp\ involving supersymmetric 
observables constructed from { \it complex matrices } 
does not follow directly 
from group theory of the unitary group, but 
rather, as the derivation shows, by the connection 
between free field contractions of the complex matrices
 and symmetric group sums followed by 
the relations between Unitary and Symmetric groups. 
Further insight into the result \fanp\ 
will be developed in the last section from 
a reduced Matrix Model of complex Matrices.

\newsec{ Three-point functions } 

We  consider the following three-point function of 
Schur polynomials. 
\eqn\theqs{ 
< \chi_{R_1} ( \Phi ) \chi_{R_2} ( \Phi ) \chi_S ( \Phi^{*} ) > }  
Representation $R_1$ has $n_{R_1}$ boxes, $R_2$ has $n_{R_2} $ and 
$S$ has $n_S$. For a non-zero free field correlator, 
$ n_S = n_{R_1} + n_{R_2} $. 
By using the expansion of Schur Polynomials in terms of characters
of the symmetric group \charp, this is 
\eqn\thptcpt{
 < \sum_{\sigma_1, \sigma_2, \tau } 
 { \chi_{R_1}  ( \sigma_1 ) \over n_{R_1} ! } { \chi_{R_2}  ( \sigma_2 )
\over n_{R_2} ! }    ~~tr ( \sigma_1 \Phi ) ~~~ tr ( \sigma_2 \Phi ) 
 ~~  { \chi_S ( \tau ) \over n_S! } ~~ tr ( \tau \Phi^{* } ) > }
where the sum over $\sigma_1$ runs over all permutations in 
$S_{\nro}$, $\sigma_2$ runs over all permutations 
in $S_{\nrt}$, and $\tau $ runs over permutations in 
$ S_{n_S}$. Expanding out the expressions of the form $tr ( \sigma
\Phi )$ using  \cptts\ we get 
\eqn\tht{\eqalign{& 
\sum_{ \sigma_1, \sigma_2, \tau } { \chi_{R_1}  ( \sigma_1 ) \over
n_{R_1} ! } 
{ \chi_{R_2}  ( \sigma_2 )
\over n_{R_2} ! }  { \chi_S ( \tau ) \over n_S ! }   \cr 
& \qquad\qquad 
< \Phi \pmatrix{ & I_1 (n_{R_1} ) \cr  
                  & I_1 ( \sigma_1 ( n_{R_1} ) )  \cr } 
      \Phi \pmatrix{ & I_2 (n_{R_2} ) \cr  
                  & I_2 ( \sigma_2 ( n_{R_2} ) ) \cr } \Phi^{*} 
 \pmatrix{ & J (n_{S} ) \cr  
                  & J (\tau  (  n_{S} ) )  \cr } > \cr }}
After doing the contractions we have a sum over 
an extra permutation $ \alpha $ in $S_{n_S} $, 
\eqn\thti{\eqalign{& 
 \sum_{ \sigma_1, \sigma_2, \tau, \alpha  } \sum_{I_1, I_2, J } 
{ \chi_{R_1}  ( \sigma_1 ) \over n_{R_1} ! } { \chi_{R_2}  ( \sigma_2 )
\over n_{R_2} ! }  { \chi_S ( \tau ) \over n_S ! }  \cr 
& \qquad\qquad\qquad\qquad
\delta \pmatrix{ & I_1( n_{R_1} ) ~~ I_2 ( n_{R_2} ) \cr  
                    & J ( \alpha \tau (n_S ) ) \cr  } 
   ~~ \delta \pmatrix{ & J ( \alpha ( n_s ) ) \cr 
                      & I_1 ( \sigma_1 (n_1) )~~ I_2 ( \sigma_2 (n_2 ))
\cr   }  \cr }}
where we have used the compact form for products 
of delta functions explained in \idpow\ and \idpowi. It is convenient 
to combine the multi-indices $(I_1,I_2) $ into a single multi-index
$I$
\eqn\thtii{\eqalign{ 
& \sum_{ \sigma_1, \sigma_2, \tau, \alpha } \sum_{I, J }
{ \chi_{R_1}  ( \sigma_1 ) \over n_{R_1} ! } { \chi_{R_2}  ( \sigma_2 )
\over n_{R_2} ! }  { \chi_S ( \tau ) \over n_S ! }  \cr 
&\qquad\qquad\qquad
 \delta \pmatrix{ & I ( n_{S} ) \cr 
                    & J ( \alpha \tau (n_S ) ) \cr  } 
    ~~~  \delta \pmatrix{& J ( \alpha ( n_s ) ) \cr 
                      & I  ( \sigma_1 \circ  \sigma_2 ( n_S )) \cr }
\cr }}
Now we do the sum over the $J$ multi-index, to get
\eqn\tt{\eqalign{&  
 \sum_{\alpha, \sigma_1, \sigma_2, \tau } \sum_{I } 
{ \chi_{R_1}  ( \sigma_1 ) \over n_{R_1} ! } 
{ \chi_{R_2}  ( \sigma_2 )
\over n_{R_2} ! }  { \chi_S ( \tau ) \over n_S ! }~~
  \delta \pmatrix{ & I ( \tau^{-1}  \alpha^{-1}( n_{S} ) ) \cr 
              & I ( \alpha^{-1} ( \sigma_1 \circ \sigma_2 ) ( n_S ) )
\cr  } \cr 
&= \sum_{\alpha, \sigma_1, \sigma_2, \tau } \sum_{I }
{ \chi_{R_1}  ( \sigma_1 ) \over n_{R_1} ! } { \chi_{R_2}  ( \sigma_2 )
\over n_{R_2} ! }  { \chi_S ( \tau ) \over n_S ! }
~~ \delta \pmatrix{ & I ( n_{S} )  \cr  
&  I( \alpha  \tau  \alpha^{-1} ( \sigma_1 \circ \sigma_2 ) ( n_S )
)\cr }     \cr }} 
We use \idpowi\ to rewrite this as 
\eqn\tti{ 
 \sum_{\alpha, \sigma_1, \sigma_2, \tau} 
{ \chi_{R_1}  ( \sigma_1 ) \over n_{R_1} ! } 
{ \chi_{R_2}  ( \sigma_2 )
\over n_{R_2} ! }  { \chi_S ( \tau ) \over n_S ! } 
  N^{C ( \alpha  \tau  \alpha^{-1} ( \sigma_1 \circ \sigma_2 ) )
}  }
Introducing an extra sum over a symmetric group 
element and constraining it by insertion of a delta function 
over the symmetric group 
\eqn\intdelt{ 
\sum_{\alpha, \sigma_1, \sigma_2, \tau, p } 
{ \chi_{R_1}  ( \sigma_1 ) \over n_{R_1} ! } 
{ \chi_{R_2}  ( \sigma_2 )
\over n_{R_2} ! }  { \chi_S ( \tau ) \over n_S ! } 
  N^{C ( p ) } 
 \delta ( p^{-1} \alpha  \tau  \alpha^{-1} ( \sigma_1 \circ \sigma_2 )
) }
Performing the sum over $\tau $
\eqn\idltf{ 
\sum_{\alpha, \sigma_1, \sigma_2, p } 
{ \chi_{R_1}  ( \sigma_1 ) \over n_{R_1} ! } 
{ \chi_{R_2}  ( \sigma_2 )
\over n_{R_2} ! }  { \chi_S ( \alpha^{-1} p ( \sigma_1^{-1} \circ
\sigma_2^{-1} ) \alpha ) \over n_S ! } 
  N^{C ( p ) } }
The sum over $\alpha$ is trivial and 
cancels the factor of $n_S!$ in the denominator
\eqn\ttii{ 
  \sum_{ \sigma_1, \sigma_2, p  } 
{ \chi_{R_1}  ( \sigma_1 ) \over n_{R_1} ! } 
{ \chi_{R_2}  ( \sigma_2 )\over n_{R_2} ! }  ~~
 \chi_S ( ( \sigma_1 \circ
\sigma_2 )^{-1} p )    N^{C(p) } }
The sum $\sum_p N^{C(p)} p$
commutes with any element of $S_{n_S}$ and therefore can be taken
out of the character $\chi_S$ using \commt\ to obtain  
\eqn\ttiii{
  \sum_{ \sigma_1, \sigma_2, p  } 
{ \chi_{R_1}  ( \sigma_1 ) \over n_{R_1} ! } 
{ \chi_{R_2}  ( \sigma_2 )
\over n_{R_2} ! }  ~~ 
{ \chi_S (     ( \sigma_1 \circ
\sigma_2 )^{-1} )  }   N^{C(p)} {\chi_S (p) \over d_S}. }
We expand the character of a permutation living 
in a product sub-group into a product of characters using \keq,  
\eqn\ti{ 
\sum_{ \sigma_1, \sigma_2, p  } 
{ \chi_{R_1}  ( \sigma_1 ) \over n_{R_1} ! } 
{ \chi_{R_2}  ( \sigma_2 )
\over n_{R_2} ! }  ~~ 
\sum_{ \rop, \rtp } g( \rop, \rtp; S ) \chi_{ \rop } ( (\sigma_1)^{-1} ) 
~~ \chi_{\rtp} ( (\sigma_2)^{-1} ) {\chi_S (p) \over d_S} N^{C(p)}. }
Finally we use orthogonality of characters  of the symmetric group \orthii\
and  \shudim\ to get the answer
\eqn\thrans{ 
g(R_1, R_2; S )
{ Dim_N ( S) ~~ n_S !  \over d_S } = g(R_1, R_2; S ) f_S. }

\newsec{ Multi-point functions }

\subsec{ Derivation of result for $(l,1)$ multi-point functions } 

When we apply the manipulations of section 5 to the
$(l,1)$ point function 
\eqn\thlon{ 
< \chi_{R_1} ( \Phi ) \chi_{R_2}  ( \Phi ) \cdots 
  \chi_{R_n}  ( \Phi  )  \chi_{S} ( \Phi^{*}  )  > } 
we are lead to an analog of \ttiii\ which is 
\eqn\anlg{ 
\sum_{\sigma_1, \sigma_2, \cdots \sigma_l, p } ~~~
 \prod_{i=1}^{l} { \chi_{R_i}  ( \sigma_1 ) \over n_{R_i} ! } 
~~~ { \chi_S (     ( \sigma_1 \circ
\sigma_2 \cdots \sigma_l   )^{-1} )  }  ~ N^{C(p)}~ {\chi_S (p) \over d_S}. }
To expand the character of a product $\chi_S (     ( \sigma_1 \circ
\sigma_2 \cdots \sigma_l   )^{-1} )  $
we make repeated use of \keq. This leads to 
\eqn\symult{ \chi_S ( \prod_i \sigma_i^{-1}  ) 
= \sum_{R_1, R_2 \cdots R_l }    \sum_{S_1, S_2 \cdots S_{l-2} } 
g(R_1, R_2; S_1) g(S_1, R_3; S_2) \cdots g(S_{l-2}, R_l ; S )  
\prod_{i} \chi_{ R_i } ( \sigma_i^{-1}  ) }  
We can recognize this, using \gmgs,  as 
\eqn\symult{ \chi_S ( \prod_i \sigma_i^{-1}  ) 
= \sum_{R_1, R_2 \cdots R_l }  g(R_1, R_2, \cdots R_l; S ) 
\prod_{i} \chi_{ R_i } ( \sigma_i^{-1}  ) } 
The remaining steps proceed as in section 5, to give the answer
\nrhto\ and \nrhtox\ after introducing the spacetime 
dependences.

\subsec{ Derivation of the $(l,k)$ multi-point function  }

The correlation function of interest is : 

\eqn\cornm{  
\sum_{\sigma_i \in S_{\nri} } \sum_{  \tau_j \in S_{\nsj}  } 
<  \prod_i { \chi_{R_i}  ( \sigma_i ) \over \nri }  ~~ tr ~( \sigma_i \Phi ) 
  \prod_j    {\chi_{S_j} ( \tau_j ) \over \nsj } ~~
 tr ~( \sigma_j \Phi^{*} ) >   }  
When we introduce the spacetime coordinate dependences, 
all the $ \Phi^{*} $ operators are at a point. 

Unravelling the traces : 

\eqn\cornmi{\eqalign{ 
&\sum_{\sigma_i \in S_{\nri} } \sum_{  \tau_j \in S_{\nsj}  } 
\prod_i { \chi_{R_i}  ( \sigma_i ) \over \nri ! } ~~  
\prod_j    {\chi_{S_j} ( \tau_j ) \over \nsj ! } \cr 
& \sum_{I_1, I_2, \cdots I_l } \sum_{J_1, J_2 \cdots J_k }  
< \Phi \pmatrix{ & I_1(n_{R_1} ) \cr 
                 & I_1( \sigma_1 (n_{R_1} ) ) } ~~
  \Phi \pmatrix{ & I_2(n_{R_2}) \cr 
                 & I_2( \sigma_2 (n_{R_2}) ) } ~~ 
  \cdots ~~   
   \Phi \pmatrix{ & I_l(n_{R_l} ) \cr 
                 & I_l( \sigma_l (n_{R_l} ) ) } ~~ \cr 
&  \qquad\qquad \Phi^{*}  \pmatrix{ & J_1(n_{S_1} ) \cr 
                 & J_1( \tau_1 (n_{S_1} ) ) } ~~
\Phi^{*}  \pmatrix{ & J_2(n_{S_2} ) \cr 
                 & J_2( \tau_2 (n_{S_2} ) ) } ~~
\cdots ~~ 
\Phi^{*}  \pmatrix{ & J_k  (n_{S_k}  ) \cr 
                 & J_k  ( \tau_k  (n_{S_k}  ) ) }  > ~~ \cr }}
We have used the multi-index notation introduced in section 2.
After we do the contractions, we get a sum over permutations
in $S_{n_T} $ where $ n_T = \sum_{i=1}^{l} \nri =  \sum_{j=1}^{k} \nsj
$. 
\eqn\perms{\eqalign{ 
& \sum_{\sigma_i \in S_{\nri} } \sum_{  \tau_j \in S_{\nsj}  } 
\prod_i { \chi_{R_i}  ( \sigma_i ) \over \nri ! } ~~  
\prod_j    {\chi_{S_j} ( \tau_j ) \over \nsj ! } \cr 
& \sum_{I} \sum_{J }  
\sum_{\alpha \in S_{n_T} } 
~~ \delta \pmatrix{ & I (n_T )  \cr 
                    & J ( \alpha  \prod_j \tau_j   ( n_T ) ) \cr } 
    \delta \pmatrix{ & J ( n_T )  \cr 
                     & I ( \alpha^{-1}  \prod_i \sigma_i   ( n_T ) )
\cr }  \cr } }
In this equation we have replaced the set of 
multi-indices $( I_1, I_2 \cdots I_k )$ by a 
single multi-index $I$, which is convenient because we 
have  permutations which mix the entire set of 
$n_T$ indices. 

Peforming the sum over the $J$ multi-index 
we are left with 

\eqn\aftri{ 
     \sum_{\sigma_i \in S_{\nri} } \sum_{  \tau_j \in S_{\nsj}  } 
\prod_i { \chi_{R_i}  ( \sigma_i ) \over \nri ! } ~~  
\prod_j    {\chi_{S_j} ( \tau_j ) \over \nsj ! } 
\sum_{I  }
\sum_{\alpha \in S_{n_T} } 
 \delta \pmatrix{  & I ( ~ \prod_j \tau_j^{-1} ~  \alpha^{-1} (n_T ) ~)\cr 
                   & I ( ~ \alpha^{-1}  ~ \prod_i \sigma_i   (n_T) ~)
     }  }

Now we use the result \idpowi\ described in 
section 2, to get : 

\eqn\aftri{\eqalign{  
&     \sum_{\sigma_i \in S_{\nri} } \sum_{  \tau_j \in S_{\nsj}  } 
\prod_i { \chi_{R_i}  ( \sigma_i ) \over \nri ! } ~~  
\prod_j    {\chi_{S_j} ( \tau_j ) \over \nsj ! } \cr 
& \sum_{\alpha \in S_{n_T} } 
 N^{ C ( \prod_i  \sigma_i^{-1}  \alpha
               \prod_j \tau_j^{-1}  \alpha^{-1}  ) } \cr }}

It will be convenient to rewrite this by introducing 
an extra sum over a permutation $p$ in $ S_{N_T} $, 
and constrain the sum by inserting a delta function over the symmetric 
group. 
\eqn\aftri{ 
     \sum_{\sigma_i \in S_{\nri} } \sum_{  \tau_j \in S_{\nsj}  } 
\prod_i { \chi_{R_i}  ( \sigma_i ) \over \nri ! } ~~  
\prod_j    {\chi_{S_j} ( \tau_j ) \over \nsj ! } 
\sum_{\alpha \in S_{n_T} } \sum_{p \in S_{n_T} } 
 N^{ C(p) } \delta ( p ~ \prod_i \sigma_i^{-1}  \alpha
               \prod_j \tau_j^{-1}      \alpha^{-1}   ) }

Now we expand the delta function into a sum over characters
using \deltaeq. 
\eqn\aftrii{\eqalign{ 
&     \sum_{\sigma_i \in S_{\nri} } \sum_{  \tau_j \in S_{\nsj}  } 
\prod_i { \chi_{R_i}  ( \sigma_i ) \over \nri ! } ~~  
\prod_j    {\chi_{S_j} ( \tau_j ) \over \nsj ! } \cr 
& \sum_{\alpha \in S_{n_T} } \sum_{p \in S_{n_T} } 
\sum_{ T \in Rep ( S_{n_T}  ) }
   { d_T \over n_T ! } 
N^{C(p)}  \chi_T ( p^{-1} ~~ \prod_i \sigma_i^{-1} ~~ \alpha
    ~~ \prod_j \tau_j^{-1} ~~    \alpha^{-1}   )   \cr }}

We can expand the character in the last line into a product of
characters
by taking advantage of the fact that $ \sum_{p} N^{C(p) } p $ 
as well as $ \sum_{\alpha } \alpha
    ~~ \prod_j \tau_j^{-1} ~~    \alpha^{-1} $ 
are in the centre of the group algebra of $S_{n_T}$, 
and the equation \commt.

\eqn\splitchar{\eqalign{& 
\sum_{\sigma_i \in S_{\nri} } \sum_{  \tau_j \in S_{\nsj}  } 
\prod_i { \chi_{R_i}  ( \sigma_i ) \over \nri ! } ~~  
\prod_j    {\chi_{S_j} ( \tau_j ) \over \nsj ! } \cr 
& \sum_{\alpha \in S_{n_T} } \sum_{p \in S_{n_T} } 
\sum_{ T \in Rep ( S_{n_T}  ) }
   { d_T \over n_T ! } 
N^{C(p)}  \chi_T ( p  )  ~~{ 1 \over d_T } 
 \chi_T ( \prod_i \sigma_i^{-1} ~~ )  { 1\over d_T } ~ 
  \chi_T ( \alpha
    ~~ \prod_j \tau_j^{-1} ~~    \alpha^{-1}   )  \cr 
& = \sum_{\sigma_i \in S_{\nri} } \sum_{  \tau_j \in S_{\nsj}  } 
\prod_i { \chi_{R_i}  ( \sigma_i ) \over \nri ! } ~~  
\prod_j    {\chi_{S_j} ( \tau_j ) \over \nsj ! } \cr 
&  \sum_{p \in S_{n_T} } 
\sum_{ T \in Rep ( S_{n_T}  ) }
   { 1 \over d_T } 
N^{C(p)}  \chi_T ( p  )  ~~
 \chi_T ( \prod_i \sigma_i^{-1} ~~ )  
  \chi_T (    ~~ \prod_j \tau_j^{-1} ) ~~         
 \cr }}

In the second line above we have done the 
sum over $\alpha$, using the invariance of the character under
conjugation, 
to get a factor of $n_T!$.  
We now apply the expansion 
\symult\ to  $\chi_T ( \prod_i \sigma_i^{-1} ~~ )$
 and   $\chi_T (    ~~ \prod_j \tau_j^{-1} )$ 
 to get a product of characters. 

\eqn\expndt{\eqalign{&
  \sum_{\sigma_i \in S_{\nri} } \sum_{  \tau_j \in S_{\nsj}  } 
\prod_i { \chi_{R_i}  ( \sigma_i ) \over \nri ! } ~~  
\prod_j    {\chi_{S_j} ( \tau_j ) \over \nsj ! } \cr 
&  \sum_{p \in S_{n_T} } 
\sum_{ T \in Rep ( S_{n_T}  ) }
   { 1 \over d_T } 
N^{C(p)}  \chi_T ( p  )  ~~ \cr 
&  \sum_{ \rop, \rtp, \dots \rlp} 
 \sum_{ \sop, \stp, \dots \skp } 
 g(\rop, \rtp, \cdots \rlp; T ) g( \sop, \stp, \dots , \skp ; T )  
  \prod_i \chi_{ \rip}  (  \sigma_i^{-1} ~~ )  ~~
   \prod_j \chi_{\sjp}  (    ~~  \tau_j^{-1} ) ~~         
 \cr }}

Now we can use orthogonality of characters to do the 
sums over permutations $ \sigma_i $, $\tau_j$, 
to obtain $ \prod_i \delta_{R_i, \rip } \prod_{j} \delta_{ S_j, \sjp }
$, which allows us to the sums over $ \rip, \sjp$ trivially.  
We also recognize the sum  $  \sum_p N^{C(p)}  \chi_T ( p  ) = n_T ! Dim
( T) $. Putting these elements together we arrive at the final formula 
for the $ ( l \rightarrow k )$ multi-point function 
\eqn\finform{ 
 \sum_{T}~~ g(R_1, R_2, \cdots , R_l ; T )
~~ { n_T! ~~ Dim_N ( T ) \over d_T } 
          ~~ g ( S_1, S_2 \cdots , S_k ; T ). } 

\newsec{ Correlators of  half-BPS
composites and gravitons in $AdS_5
\times S^5$  } 

Young Diagrams $R$ associated with  small $\car$-charge, 
i.e $ n_R \ll N$ are associated with KK states. 
They can be written as sums of products 
of small numbers of traces.  
There is an approximate correspondence in the large N limit
between  single trace chiral primaries and single particle states
in AdS via the AdS/CFT correspondence. Kaluza-Klein
modes arising from dimensional reduction of ten-dimensional
type IIB supergravity on $S^5$ couple to single trace chiral
primaries on the boundary of $AdS_5$.   
Using the map given in  \gkp\withol, 
one can compute correlators of the KK modes and compare
to correlators of the single trace chiral primaries and
one finds agreement in the large N limit for the
three point functions, \refs{\lmrs}.

Among Young Diagrams  with large numbers 
of boxes, two classes stand out. 
One involves large mostly-antisymmetric representations 
and another involves large mostly-symmetric representations. 
Among the large, mostly antisymmetric ones, there is a 
series of Young Diagrams 
which have one column of length $L$, 
where $L$ is comparable to $N$ and the remaining 
columns of length zero. We may denote these
by the column lengths $ \vec c = ( L, \vec 0 ) $. 
These 
are proportional  to the sub-determinants considered 
in \bbns\ as duals of sphere giants
\eqn\subdl{ \chi_{(L, \vec 0 ) } ( \Phi ) \sim 
 {1 \over L!} \epsilon_{i_1 \cdots
i_L i_{L +1} \cdots i_N } \epsilon^{j_1 \cdots j_L j_{+1}
\cdots j_N} \Phi^{i_1}_{j_1} \cdots \Phi^{i_L}_{j_L} }
This series has a cutoff of $L \le N$ which 
 corresponds to the fact that sphere giants 
have a maximal angular momentum of $N$. 

More general representations among the mostly-antisymmetric 
ones have column lengths $ \vec c = ( L_1, L_2, \cdots L_k, \vec 0 )$. 
All the $L_1 \cdots L_k$ are comparable to $N$ 
and the number of non-zero column lengths $k$ is small. 
If all of them are equal, a natural dual giant 
 graviton is a simple generalization of the original 
sphere giant, where the three-brane wraps the 
$S^3$ in $S^5$ with a winding number $k$. 
An ansatz with winding number $k$ leads to 
 an effective probe Lagrangian of the same form 
as that for the original sphere giant, with the factor 
$k$ scaling out of the Lagrangian. This leads to 
a solution with angular momentum and energy scaled up 
by a factor of $k$. When the $L_i$ are not all equal, 
but differ by numbers small compared to $N$,  
we can think of the Young Diagram as obtained by
fusing a large Young Diagram with all equal and large 
columns with a small Young Diagram. Thus the natural 
physical interpretation is in terms of composites 
involving giants and Kaluza-Klein gravitons. 

The simplest large mostly-symmetric representations 
have one row of non-zero length $L$, 
so the vector $ \vec r $ of row lengths 
is $ \vec r = ( L, \vec 0) $. Now $L$ has no
upper bound and this series has angular momenta 
increasing in units of $1$. These properties 
suggest that these are duals of AdS giants. 
Indeed AdS giants have no upper bound on their 
angular momentum \refs{\gmt,\hashitz, \djri }. 
Duals to AdS giants at the classical level 
have been proposed in \hashitz. We find there is indeed
a heuristic  connection between the classical solutions 
considered there and these symmetric representations. 
The solutions in \hashitz\ have the form : 
\eqn\frmsol{ \Phi \sim  \Phi_0 e^{ i \omega t } } 
where $ \Phi_0 \sim  Diag ( \eta, -{ \eta \over N-1} \cdots - { \eta \over
N-1} )$. Viewing $ \chi_R ( \Phi )$ as wavefunctions 
we would expect a strong overlap between the candidate
dual to the AdS giant and the corresponding solution. 
Indeed, approximating at large $N$ the matrix $ \Phi_0$
with $ Diag ( \eta, \vec 0 )$, we find that 
$ \chi_R ( \Phi_0 ) $ is zero unless $R$ is the symmetric 
representation with one non-zero row length. 

For these reasons, when $R$ is described 
by the vector of row lengths $ \vec r = ( L, \vec 0) $, 
with $L$ comparable to or much larger than $N$, 
we propose the spacetime dual to be 
the AdS giant. For $ \vec r = ( L,L, \cdots L , 0 )$ 
with a small number $k$ of non-zero 
entries, we associate it with a multiply 
wound AdS giant, i.e where the 3-brane multiply 
winds the $S^3$ in $AdS_5$. As in the sphere giant 
case, the multiple winding ansatz leads to an 
effective action which has the winding number $k$  
scaling out, leading to a scaling of the angular 
momentum and energy by $k$ in agreement with the 
proposed gauge theory dual. 
As in the discussion of sphere giants, the 
Young Diagram with row lengths 
$ \vec r = ( L_1, L_2, \cdots, L_k , \vec 0 )$
where the $L_i$ are not all equal, can be considered 
to be duals of composite  states 
involving  AdS giants and KK gravitons. 
 
 The detailed test of this picture will 
 emerge by comparison of correlators 
 of giants computed in gauge theory 
 and computations in string theory 
 on $AdS_5 \times S^5$. We will write 
 down  gauge theory correlators
 involving the various types 
 of gravitons, applying the general formulae derived 
 in previous sections. We leave the computations 
 on the string theory side to future work. Even a clear formulation 
 of semiclassical gravity + probe calculations 
 of these correlators remains to be articulated.
 Our remarks on the interpretation of the large $N$ behaviour 
 of the correlators we describe below 
 are therefore somewhat heuristic.

\subsec{ Large antisymmetric representations  and Sphere Giants } 

For the correlators that we shall consider in this and the
next sections, we shall discuss two natural ways in which
to normalize.  For a correlator of the form
$\langle \prod_{i=1}^{l} \chi_{R_i}(\Phi) \prod_{j=1}^{k} 
\chi_{S_j}(\Phi^{*}) \rangle$, we could normalize it
by dividing by the norms of each individual character,
i.e., by $\prod_{i=1}^{l} \parallel \chi_{R_i}(\Phi) \parallel$ and
$\prod_{j=1}^{k} \parallel \chi_{S_j}(\Phi) \parallel$. 
Another possibility is to divide instead by the norm of the
$\Phi$ operator 
$\parallel \prod_{i=1}^{l} \chi_{R_i}(\Phi) \parallel$ as well
as by the norm of the $\Phi^{*}$ operator
$\parallel \prod_{j=1}^{k} \chi_{S_j}(\Phi^{*}) \parallel$. 
The latter normalization is relevant if the correlator is
to be interpreted as an overlap of the two states
$\prod_{i=1}^{l} \chi_{R_i}(\Phi^{*}) |0 \rangle$ and
$\prod_{j=1}^{k} \chi_{S_j}(\Phi^{*}) |0 \rangle$. 
In this case one can show using \nrhtm\ that the correlator
is bounded above by one.  As a simple illustration consider
the $m=1$ case.  From \nrhtm\ we have
\eqn\normeg{
\parallel \prod_{i=1}^{l} \chi_{R_i}(\Phi) \parallel =
\sqrt{\sum_T (g(R_1,R_2,...,R_n;T))^2
{n_{T}! ~ dim_{N}T \over d_T}}} 
while $\parallel \chi_S(\Phi) \parallel = \sqrt{n_{S}! ~ dim_{N} S/d_S}$.
Dividing \nrhto\ by these two norms gives a normalized
correlator which is less than one, since the 
sum over $T$ always contains $S$ in addition to additional 
positive terms.  The general case follows
in a similar manner.

Now we turn to some explicit examples of correlators
involving large representations, i.e., representations
with ${\cal O}(N)$ boxes in the associated Young Diagrams. 
As a first example we consider the three point
correlators $\langle \chi_{L_1}(\Phi) \chi_{L_2}(\Phi) 
\chi_{L_3} (\Phi^{*}) \rangle$ where $L_3 = L_1 + L_2$ 
and the representations
are the completely antisymmetric representations derived
from the $L_{i}^{th}$ tensor product
representation. 
From \threept\fanp\fr\  we can read off the
normalized three point function as
\eqn\antithree{{\langle \chi_{L_1}(\Phi) \chi_{L_2}(\Phi) 
\chi_{L_3} (\Phi^{*}) \rangle \over 
\parallel \chi_{L_1}(\Phi) \parallel
\parallel \chi_{L_2}(\Phi) \parallel
\parallel \chi_{L_3}(\Phi) \parallel}
= \sqrt{{N! \over (N-L_3)!}{(N-L_1)! \over N!}{(N-L_2)! \over N!}.}}
We have used the fact that the Littlewood-Richardson 
coefficient in this case is $1$, as can be deduced 
from the rules in \baracrule\ for example.  

Taking in particular $L_3 = N$ with $L_1 = (N/2 + L)$ and
$L_2 = (N/2 - L)$ we obtain
\eqn\antithreeN{{{\langle \chi_{(N/2 -L)}(\Phi) \chi_{(N/2 + L)}(\Phi) 
\chi_N (\Phi^{*}) \rangle \over \parallel \chi_{(N/2 +L)}(\Phi)
\parallel \parallel \chi_{(N/2 - L)}(\Phi) \parallel
\parallel \chi_{N}(\Phi) \parallel}}
= \sqrt{{(N/2+L)! (N/2-L)! \over N!}}.}  For $L = {\cal O}(1)$ this 
correlator represents the three point function of sphere
giant gravitons, considered in \bbns. 
For $L=0$ the correlator is approximately
$1/2^{N/2}$.  This exponential decay in $N$ suggests that
an instanton mediates a transition between the single sphere giant
with R-charge $N$ and the pair of giants of R-charges $N/2$.    
The right-hand-side of \antithreeN\ monotonically increases
with $L$ until at $L=(N/2 -1)$ we get a correlator going like $1/N$.
So in fact for all $0 \leq L \leq (N/2 - 1)$ the correlator is 
either exponentially or polynomially suppressed in $N$.

To normalize the correlator \antithree\ as an overlap
of two states as discussed above requires computing
a four point function.  This follows straightforwardly
from \nrhtm\ and the result is 
\eqn\fourpt{\parallel \chi_{L_1}(\Phi) \chi_{L_2}(\Phi) \parallel^{2}
= \sum_{l=0}^{L_2} {N! (N+1)! \over (N-L_1 -l)! (N-L_2 + l+1)!}.}
We have used the fact that the 
tensor product of the representation 
with column lengths $ ( L_1, \vec 0 ) $ with 
$(L_2, \vec 0 ) $ ( for $ L_2  < L_1 $ ) 
 contains with unit multiplicity
the representations $ ( L_1 + l , L_2 - l , \vec 0 ) $
with unit multiplicity, with $k$ ranging from 
$ 0$ to $L_2$.    
For the special case of $L_1 = L_2 = N/2$ the sum can be
done and one finds the very simple expression $2^N N!$.
Normalizing the three-point function \antithreeN\ for
$L=0$ in this way leads to the exact answer of
\eqn\antithreeexact{{{\langle \chi_{N/2}(\Phi) \chi_{N/2}(\Phi) 
\chi_N (\Phi^{*}) \rangle \over \parallel (\chi_{N/2}(\Phi))^2
\parallel
\parallel \chi_{N}(\Phi) \parallel}}
= {1 \over 2^{N/2}}.}  So even with this normalization we
find an exponential decay in $N$ of the overlap.

Another case of interest is the three point function of
characters in the representations as in \antithree\ with
the $L_3$ representation replaced by a Young tableaux
with two columns, one of length $L_1$ and the other of
length $L_2$.  One finds the the simple looking result
\eqn\multiwrap{{\langle \chi_{L_1}(\Phi) \chi_{L_2}(\Phi) 
\chi_{L_1,L_2} (\Phi^{*}) \rangle \over 
\parallel \chi_{L_1}(\Phi) \parallel
\parallel \chi_{L_2}(\Phi) \parallel
\parallel \chi_{L_1,L_2}(\Phi) \parallel}
= \sqrt{N+1 \over N-L_2 +1}} 
assuming that $L_2 \leq L_1$.  As $L_2$ ranges from 1 to $L_1$
one finds that the correlator either remains ${\cal O}(1)$ or
becomes ${\cal O}(\sqrt{N})$ if $L_1 \approx N$. 
 For $L_1$ and $L_2$ both small, the three point functions 
 involves KK states rather than giant gravitons.  
 The Schur Polynomials are sums of traces and multi-traces 
 with coefficients of ${\cal O}(1)$. The single traces 
 are associated with single particle states in the leading 
 large $N$ limit and they have three point functions of 
 which go like  ${\cal O}({1\over N}   )$. The generic Schur polynomials 
 have $  {\cal O}(1)$ correlators because there are contributions 
 from disconnected diagrams.  
 For $L_2 =
{\cal O}(1)$ with $ L_1 = { \cal O} ( N)$ 
this correlator may be associated with  a sphere
giant and a multi-particle Kaluza-Klein state. 
The fact that the correlator is $ { \cal O} ( 1) $
might be interpreted as predicting that the KK state 
and the giant graviton do not form a bound state
which would have appeared as a half-BPS fluctuation 
around the sphere giant 
(which has not been found in the literature). 
This statement should be interpreted with care 
since we do not fully understand the rules 
for computing correlators of giants in the spacetime 
approach.   

In the opposite limit, $L_2 = L_1$, the result
that the correlator is at least ${\cal O}(1)$ 
or actually growing with $N$ for large $L_1$
is somewhat puzzling given our earlier argument
that there should exist multiple-wrapped sphere
giant gravitons.  From a supergravity perspective
however this probably should not be too suprising
as these giants would all be on top of one another
and therefore would overlap significantly.  
It is interesting to note that while the
correlator that we are discussing does not 
have a nice large $N$ limit, it does have 
a nice small $N$ limit ! 

Had we normalized the correlator \multiwrap\ using
the overlap of states prescription then our 
previous arguments would have ruled out the
growth in $N$ that we found above.  In fact for
$L_1 = L_2 = N/2$ we find that the correlator
decays as $1/N^{1/4}$ as opposed to the ${\cal O}(1)$
behaviour found above.  Furthermore for $L_1 = L_2 =N$
we find that the correlator is $1/\sqrt{1 + 1/N} \approx 1$
as opposed to the $\sqrt{N}$ growth found in the
other normalization.

Our last example is the overlap between a sphere
giant and a multi-particle Kaluza-Klein state,
\eqn\multitoone{{\langle (\chi_{1}(\Phi))^L 
\chi_{L} (\Phi^{*}) \rangle \over 
\parallel (\chi_{1}(\Phi))^L \parallel
\parallel \chi_{L}(\Phi) \parallel}
= \sqrt{{1 \over N^L}{N! \over L! (N-L)!}}.}
As before the subscript
$L$ denotes the totally antisymmetric representation
and the subscript $1$ the fundamental.  This
clearly decays exponentially for large $L$, 
which may be used as a prediction for the existence 
of a semiclassical giant graviton as a distinct 
object from multi-particle KK state.

\subsec{ Large Symmetric representations and AdS  giants }
 
Let us  now consider some examples of correlators of
characters in large symmetric representations, 
which we argued are duals of AdS giants. 
The first thing to consider is the 
large $ N$ behaviour of the 
overlap between a large number of single trace operators 
and a symmetric representation 
\eqn\ovrlap{ \langle (tr ( \Phi ))^L \chi_{L } ( \Phi^{* }  )
\rangle \over \parallel ( tr (\Phi))^L \parallel 
               \parallel \chi_{L}( \Phi^{*}  \parallel  )  }  
In this section we are using $\chi_{L }$ to denote
 the character in the representation with one 
row of length $L$. 
The norm  $\parallel ( tr (\Phi))^L \parallel$ is easily seen 
to be $ \sqrt {N^L L!}  $. The above normalized correlator 
is then found to be : 
\eqn\ansab{ \sqrt { f_L \over N^L L! } =  { ( N + L -1 )! \over ( N-1
)! L! N^L  } } 
For $ N = L $ this is equal to $ (2N-1)! \over N! N^N (N-1) ! $. 
This shows a rapid decay at large $ N$ going like $ N^{-N}$. 
For $ L \gg N$ we continue to get rapid decay. 
 This  tells us that the overlap between 
 a large number of KK states and the proposed dual 
 to the AdS giant is very small in the large $N$ limit, 
 so we may expect the semiclassical description of 
 string theory on $ AdS \times S$ with gravity coupled to 
 branes to produce a dual to $ \chi_{L}( \Phi )$  
 other than the naive large $N$ extrapolation of 
 multiparticle KK states. This is the semiclassical 
 giant AdS graviton \mst\hashitz\djri. 
  
As in the case of the sphere giants, the following 
 normalized three-point function is of interest 
\eqn\symmthree{ {\langle \chi_{L_1}(\Phi) \chi_{L_2}(\Phi) 
\chi_{L_3} (\Phi^{*}) \rangle \over 
\parallel \chi_{L_1}(\Phi) \parallel
\parallel \chi_{L_2}(\Phi) \parallel
\parallel \chi_{L_3}(\Phi) \parallel}
= \sqrt{{(N+L_3 -1)! \over (N-1)!}{(N-1)! \over (N+L_1 -1)!}
{(N-1)! \over (N+L_2 -1)!}}} where $L_3 = L_1 + L_2$ and
now all characters are in the symmetric representations.
This looks roughly like the
antisymmetric representations case \antithree\ but
differs in an essential way in that the $L_i$'s now
enter inside the factorials with plus signs instead
of minus signs.  This is one sign that there is no
bound on the R-charge.  This also gives rise to
a significant difference in the behaviour of the
correlator from \antithree.  Take for example
$L_1, L_2 = N/2$, then we find
after expanding the factorials
\eqn\symmthree{{\langle (\chi_{N/2}(\Phi))^2 
\chi_{N} (\Phi^{*}) \rangle \over 
\parallel \chi_{N/2}(\Phi) \parallel^{2}
\parallel \chi_{N}(\Phi) \parallel}
= ({32 \over 27})^{N/2},} i.e., exponential
growth!  In fact it is easy to show that
this correlator grows with $L_1$ and $L_2$
when these two parameters are large, ${\cal O}(N)$
or larger say.

From our earlier general arguments we know that
this exponential growth of the correlator in $N$
would be removed in the overlap of states normalization.
While the norm $\parallel \chi_{L_1}(\Phi) \chi_{L_2}(\Phi) \parallel$
can be evaluated and expressed as a sum analogous to 
the antisymmetric case of \fourpt.

\newsec{ The Complex Matrix Model }

Consider the reduced action \hashitz. 
\eqn\redlag{ { R^3 \Omega_3 \over 2 g_{YM}^2 } 
\int dt Tr (  {\dot{ \Phi_1}}^2  +  {\dot{ \Phi_2}}^2 - 
 {1 \over R^2} ( \Phi_1^2 + \Phi_2^2 )  ) } 
Here $\Omega_3$ is the volume of $S_3$ at the boundary of 
 $AdS_5$, where the metric is : 
\eqn\metads{ 
ds_{AdS}^2 = { R^2 \over cos^2( \rho) } ( - dt^2 + d {\rho}^2 + sin^2
( \rho) d\Omega_3 ) } 

 This leads, up to an overall factor, to the Hamiltonian 
\eqn\ham{ H = { Tr  \over 2 } ( P_1^2 + P_2^2 + \Phi_1^2 + \Phi_2^2 )
} 
with the angular momentum operator given by 
\eqn\ang{ J =  Tr ( P_1 \Phi_2 - P_2 \Phi_1 ) } 
and canonical commutation relations 
\eqn\cancom{ [ (P_1)_{ij}, (\Phi_1)_{kl}  ] = 
[ (P_2)_{ij} , (\Phi_2)_{kl} ] = -i \hbar \delta_{jk} \delta_{il}  } 
 We introduce complex matrices
\eqn\compmat{\eqalign{
& Z = { 1 \over \sqrt 2 } ( \Phi_1 + i \Phi_2 ) \cr 
 &  Z^{\dagger} = { 1 \over \sqrt 2 }   ( \Phi_1 - i \Phi_2 ) \cr 
}}
and the conjugates 
\eqn\conj{\eqalign{ 
& \Pi =   { 1 \over \sqrt 2 } ( P_1 + i P_2 ) = -i 
{  \partial \over \partial Z^{\dagger} } \cr 
& \Pi^{\dagger} =  { 1 \over \sqrt 2 } ( P_1 - i P_2 ) = -i  {  \partial
\over \partial Z } \cr }}
 Since these obey relations, 
\eqn\rels{\eqalign{ 
& [ Z, \Pi ] = [ Z^{\dagger} , \Pi^{\dagger} ] =0 \cr 
& [ Z_{ij} , \Pi^{\dagger}_{kl}  ] = [ Z^{\dagger}_{ij} , \Pi_{kl}  ] 
 = i \hbar \delta_{jk} \delta_{il} \cr }}
we can define creation-annihilation operator pairs 
\eqn\crtn{\eqalign{
&   A =  { 1 \over \sqrt 2 } ( Z + i \Pi ) \cr 
&   A^{\dagger} = { 1 \over \sqrt 2 } ( Z^{\dagger}  - i \Pi^{\dagger}
) \cr }}
which obey the standard relation $ [ A_{ij} , A^{\dagger}_{kl}  ] = \delta_{jk} \delta_{il}  $. 
An independent set of creation-annihilation operator pairs can be defined, 
\eqn\ncrtn{\eqalign{
&   B =  { 1 \over \sqrt 2 } ( Z - i \Pi ) \cr 
&   B^{\dagger} = { 1 \over \sqrt 2 } ( Z^{\dagger}  + i \Pi^{\dagger}
) \cr }} 
which obey $ [ B, B^{\dagger} ] =1 $. 
It is also easily verified that 
\eqn\abrel{\eqalign{ 
& [A,B ] ~~= ~~[A, B^{\dagger} ]~~ = ~~ 0 \cr 
&  [ A^{\dagger} , B ]~~ = ~~[A^{\dagger} ,
B^{\dagger} ] ~~= 0 ~~ \cr }}

In these variables, the Hamiltonian and angular momentum are  
\eqn\haminv{\eqalign{ & 
 H = Tr ( A^{\dagger} A + B^{\dagger} B  ) \cr 
& J = Tr ( A^{\dagger} A - B^{\dagger} B ) \cr }} 

Some eigenstates of the Hamiltonian, with their 
energies and momenta are listed below
\eqn\eigs{\eqalign{ 
 Tr  (( A^{\dagger} )^n ) | 0 > ~~~ & E = J = n   \cr 
  Tr ( ( B^{\dagger} )^m )  |0 > ~~~ & E= -J = m \cr 
  Tr ( (A^{\dagger} )^n ( B^{\dagger} )^m ) |0> ~~  & E = n+m, ~ J= n-m \cr}}
For chiral primaries with $E = |J| $ we have $n=0$ or $m=0$. 

It is useful to diagonalize $A,A^{\dagger}$ by using the unitary
symmetry,  
\eqn\diagaa{\eqalign{  
& A_{ij} = \lambda_i \delta_{ij} \cr 
&  A^{\dagger}_{ij} = \lambda_{i}^{\dagger}  \delta_{ij} \cr }}
The measure, in terms of these variables, shows that 
we can treat the $\lambda_i$'s as fermionic variables.  
The Hamiltonian for these fermionic oscillators is  
\eqn\ham{ H = \sum_{i} \lambda_i^{\dagger} \lambda_i } 
The fermionic wavefunctions are 
\eqn\ferms{ \psi_F ( \lambda_1, \lambda_2, \cdots \lambda_n ) 
= e^{-\sum_i { \bar \lambda_i} \lambda_i } 
 Det \pmatrix{ 
& \lambda_1^{n_1} & \lambda_1^{n_2} & \cdots & \lambda_1^{n_N} \cr   
& \lambda_2^{n_1} & \lambda_2^{n_2} & \cdots & \lambda_2^{n_N} \cr 
& \vdots & \vdots & \vdots & \vdots \cr 
& \lambda_{N}^{n_1} & \lambda_N^{n_2}& \cdots & \lambda_{N}^{n_N} \cr }
} 
For these states the energy and angular momentum are 
$E=J = \sum_{i} n_i $. 
The ground state corresponds to 
$n_1 = N-1, n_2 =N-2 , \cdots n_N =0 $. 
The ground state  wavefunction  is a Van der Monde determinant
\eqn\gdst{ \Psi_0 = e^{- \sum_i {\bar \lambda_i}  \lambda_i } \prod_{i
< j } ( \lambda_i - \lambda_j ) }
General excited states are described by Young diagrams 
with row lengths $ \vec r = ( r_1, r_2 \cdots r_N )$. 
The energy and angular momentum are  
\eqn\energy{  E =J=  \sum_{i} r_i  + i-1 } 
The wavefunctions normalized by the Van der Monde 
can be recognized as Weyl's character formula and hence 
are the Schur Polynomials. 

\subsec{ Hints on integrability } 
 
 From \energy\ it is clear that all the states associated 
 with Young Diagrams having $n$ boxes have the same energy. 
 As such there is a large degeneracy equal to the number of 
 partitions of $n$. In \fanp\ we found that the exact two-point
 functions at finite $N$ are orthogonalized by the 
 Schur Polynomials. This suggests that there are higher 
 Hamiltonians which commute with \ham\ and which are diagonalized by
 the Schur polynomials and which have different 
 eigenvalues for different Young diagrams. 
 These higher Hamiltonians can be constructed in terms 
 of $tr ( (AA^{\dagger})^n ) $ and  $tr ( (BB^{\dagger})^n ) $.
 The $4$ dimensional origin of these is a very intereating 
 question. Presumably they will involve modifications 
 of $N=4$ Yang Mills with higher derivatives. These 
 may be expected to be unrenormalizable, but perhaps
 they have a higher dimensional origin, generalizing the 
 connection between $N=4$  SYM and 
$(0,2)$ six-dimensional superconformal theory  or little string 
theory. \refs{ \dvv, \witten, \brs, \seib } 

\newsec{ Summary and Outlook } 

 We described a one-one mapping
 between  the space of half-BPS representations
 to the space of $U(N)$ Young Diagrams, with 
 holomorphic functions  of one complex matrix 
 playing an important role in the mapping.  
 Using this map we  identified a basis 
 of composites which diagonalizes 
 the two-point functions at finite $N$. 
 The basis suggests natural candidates 
 for giant gravitons, more precisely for 
 sphere giants, AdS giants and composites 
 of giants with Kaluza-Klein states. 
 
 We computed the normalized two and three 
 point correlators of these observables, 
 as well as a special class of higher point functions. 
 The Frobenius-Schur duality  between symmetric 
 and Unitary groups played a useful role. 
 It allowed  us to use various results 
 from $U(N)$ group theory to give exact 
 answers for correlators, which could be developed 
 in a large $N$ expansion. Many of these correlators 
 have simple interpretations in terms of 
 giant sphere and AdS gravitons, multiply 
 wound giants and composites of giants with KK states. 
 However  detailed interpretations will have 
 to await a clear formulation of rules for 
 computing interactions of giants from string 
 theory in $AdS \times S $ or its semiclassical limit. 

 The holomorphic observables we considered 
 were shown to have a distinguished 
 role in a reduced $0+1$ Matrix Model
 related to $N=4$ SYM on a three sphere. 
 The discussion of the Matrix Model also
 suggested hints of integrable structure
 in the $N=4$ theory. 

 Several generalizations of this work can be contemplated. 
 A lot of information about correlators
 of descendants  at finite $N$ should 
 be extracted. Most of our formulae 
 involved dimensions and fusion coefficients 
 of Unitary groups. When the descendants 
 are involved there should be a nice generalization 
 involving group theoretic quantities associated 
 with $U(N)$ as well as the superconformal symmetry 
 group. A generalization to maximally supersymmetric 
 gauge theory with $O(N)$ and $Sp(N)$ gauge groups 
 should also be possible.

 Quarter-BPS operators, considered 
 in detail recently \refs{ \dh, \rhy }  are also of
 interest in the context of giant gravitons \mikh.  
 Many of the  symmetric group 
 techniques used here should be useful, 
 but the final answers are  unlikely to have 
 simple relations to $U(N)$, since the observables 
 of \refs{ \dh, \rhy} do not appear to be related 
 to holomorphic gauge invariant functions of a complex Matrix.  
 Whether another 
 less obvious  group or algebra plays an analogous role 
 is a fascinating question.

\bigskip

\noindent{\bf Acknowledgements:}
 We wish to thank for pleasant discussions 
 David Lowe and Andrei Mikhailov. 
 This research was supported  by DOE grant  
 DE-FG02/19ER40688-(Task A).

\listrefs

\end